\documentclass[prl,twocolumn,aps,floatfix,showpacs,tightenlines,superscriptaddress,amsmath,amssymb]{revtex4}
\usepackage{amssymb,amsmath,amsthm,amsbsy,epsfig,color,graphicx}
\usepackage[ansinew]{inputenc}
 \usepackage[english]{babel}
 \usepackage{braket}
 \usepackage{ulem}

\begin{document}

\title{Entanglement, holonomic constraints, and the quantization of fundamental interactions}

\author{Salvatore Marco Giampaolo}
\affiliation{Division of Theoretical Physics, Ru\dj{}er Bo\v{s}kovi\'{c} Institute, Bijen\u{c}ka cesta 54, 10000 Zagreb, Croatia}

\author{Tommaso Macr\`i}
\affiliation{Departamento de F\'isica Te\'orica e Experimental and International Institute of Physics,
Universidade Federal do Rio Grande do Norte, 59072-970 Natal-RN, Brazil}

\date{\today}

\begin{abstract}
It is a general belief that all fundamental interactions need to be quantized.
However, all attempts to develop a quantum theory of gravity presented various problems, leading to a recent active debate about how to probe 
its quantum nature.
In the present work we provide a proof for the necessity of quantizing fundamental interactions demonstrating that a quantum version is needed for any non 
trivial conservative interaction whose strength is a function of the relative distance between two objects.
Our proof is based on a consistency argument that in the presence of a classical field two interacting objects in a separable state could not develop 
entanglement. 
This requirement can be cast in the form of a holonomic constraint that cannot be satisfied by generic interparticle potentials. 
Extending this picture of local holonomic constraints, we design a protocol that allows to measure the terms of a multipole expansion of the interaction of two composite bodies.
The results presented in this work can pave the way for a study of fundamental interactions based on the analysis of entanglement properties.
\end{abstract}

\maketitle

More than a century after its birth, quantum mechanics is considered one of the physical theories that received greatest experimental confirmation. 
It presents several aspects such as its intrinsically non-deterministic nature as well as quantum entanglement, that do not have classical 
counterpart~\cite{Bennet2000,Nielsen2000}.
Thus, to provide a coherent picture of empirical evidence, great efforts were made to develop quantum versions of known classical theories, as for 
electromagnetism.
Nowadays, electroweak and strong interactions, each one characterized by its own gauge group and force mediators, 
are elegantly described within the Standard Model of particle physics~\cite{Griffiths1987,Greiner2000,Cohen1997,Perkins2000}.

Differently from the other forces, the quantization process of the gravity still presents several problems.
Even if different theories were developed~\cite{Kibble1979}, including loop quantum gravity~\cite{Rovelli2004,Hamma2018} and string 
theory~\cite{Polchinski2005}, all of them are affected by some kind of problems~\cite{Rothman2006,Dyson2012,Kiefer2013} and, up to now, a widely accepted 
quantum theory of gravity is still missing.
A large debate rose about the fact that a quantum version of gravity is really needed or if, on the contrary, it must be considered an 
intrinsically classic interaction~\cite{Page1981,Penrose1996,Albers2008,Boughn2009,Dyson2013,Carlip2008}.
Over the last year this discussion has been brightened by the publication of two independent works~\cite{Bose2017,Marletto2017}.
The authors of these papers suggest to use entanglement~\cite{Nielsen2000,Preskill1999}, a quantity that is playing an ever-increasing role in very different 
fields ranging from metrology~\cite{Kimble2001,Macri2016} to quantum many-body systems~\cite{Giampaolo2007,Amico2008,Giampaolo2014,Yang2017}, to test the 
quantum nature of gravity experimentally.
The key point of their proposal is that, if an interaction is intrinsically classical, and hence its evolution can be described by a 
Koopman-type dynamics~\cite{Koopman1931}, it would not be able to generate entanglement between two massive objects.
Indeed, from the point of view of quantum information theory, a classical field is equivalent to a classical channel that is unable to increase the entanglement
between two system at its end-points~\cite{Bennet1999}.   
Therefore any evidence of entanglement between the two objects, generated by gravity, would be the smoking gun of its quantum nature. 
Such entanglement does not provide any information about the right quantum theory. 
However it would prove that quantization of gravity is necessary to explain consistently the whole phenomenology associated with the 
experiments~\cite{Peres2001,Terno2006,Hall2018}.

In this letter, we present a general argument that allows us to prove that theories aiming to describe any fundamental interaction, 
including gravity, need to be quantized. 
We use the fact that the effective interparticle potential of a conservative interaction depends on the relative distance between two objects but not on the 
derivative of any order of such a distance with respect to time.
We show that the request that it does not create entanglement is equivalent to impose a holonomic constraint that cannot be satisfied for a generic setup.
To enforce the consistency of the theory, we are then led to conclude that such interaction presents a quantum nature.
Inspired by the discussion of the holonomic constraint, we propose a generalized spin-echo scheme~\cite{Bodenhausen1969} to measure subleading terms in a 
multipole expansion of the interparticle potential. 
This interferometric measurement is based on the efficient suppression of the phases of the dominant terms.

\begin{figure}[t]
\begin{center}
\includegraphics[width=1.0\columnwidth]{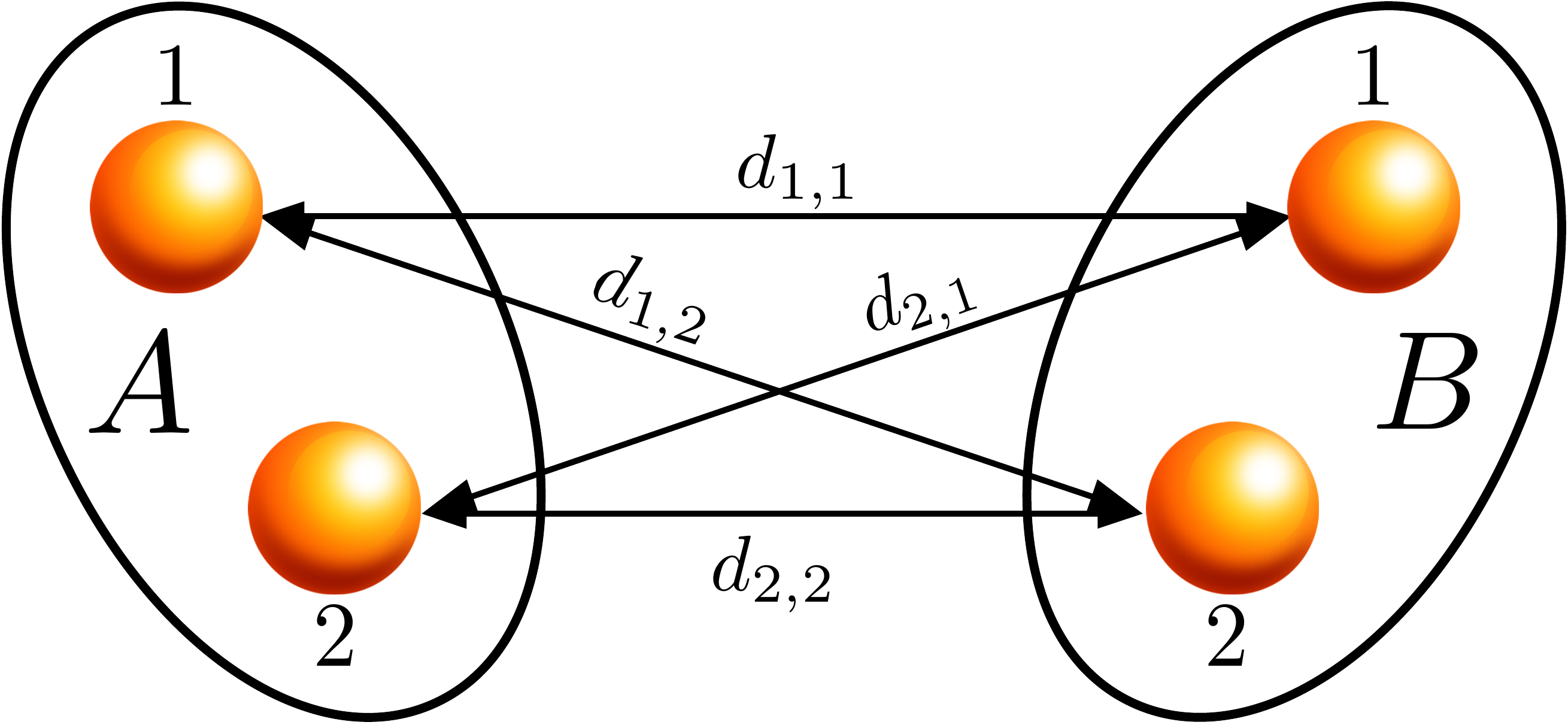}
\caption{Sketch of the scheme used in the letter. 
Two systems $A$ and $B$, represented by the two ellipses, can be in two spatially separate states, named $1$ and $2$ represented by two gray circles. 
The interaction between the two systems is function of the distance, and since it depends on the states occupied by the two systems, the interaction also 
becomes dependent on the state of the composite system.}
\label{figure_1}
\end{center}
\end{figure}

The physical system which we consider in work is represented schematically in Fig.~\ref{figure_1}.
Two physical objects, named $A$ and $B$ can be found in two different spatially separated internal states labeled $1$ and $2$.
We assume that the interaction between the two bodies is mediated, over the distance, by a conservative force field whose classical or quantum nature is the subject 
of our analysis.
Differently from all others physical characteristics, such as mass, charge, flavor etc. that we assume time-independent, the relative distances between 
the different states of $A$ and $B$ can be a function of time.
The reason to consider spatially separated states 
is to create a state-dependent interaction. 
More general implementations of potentials depending on the 
internal states will be considered below.

For the setup of Fig.~\ref{figure_1} 
the Hamiltonian of the system reads
\begin{eqnarray}
 \label{Hamiltonian}
 \hat H\!&\! =\! &\!f(d_{1,1}(t))\! \ket{1,1}\!\!\bra{1,1}+f(d_{1,2}(t)) \! \ket{1,2}\!\!\bra{1,2} \nonumber \\ 
  \!&\! + \!& \!f(d_{2,1}(t))\! \ket{2,1}\!\!\bra{2,1}+f(d_{2,2}(t))\! \ket{2,2}\!\!\bra{2,2} \, .
\end{eqnarray}
Here $f$ stands for the interaction energy that is a real function of the distance $d_{\alpha,\beta}(t)$ that depends on the state $\ket{\alpha,\beta}$ of the 
first $A$ ($\alpha$) and of the second $B$ ($\beta$) object as well as on time $t$. 
Defining, for each single object, the operators \mbox{$\sigma^z_i=\ket{1_i}\!\!\bra{1_i}-\ket{2_i}\!\!\bra{2_i}$} we have that the Hamiltonian in 
eq.~(\ref{Hamiltonian}) can be rewritten as
\begin{eqnarray}
\label{Hamiltonian1}
\hat H\!&\!\! =\!\! &\! g_1 \mathbf{1}_A\!\otimes\! \mathbf{1}_B\!+
\! g_2 \sigma^z_A\!\otimes\!\mathbf{1}_B \!+\!g_3 \mathbf{1}_A\!\otimes\!\sigma^z_B 
\! +\! g_4 \sigma^z_A \!\otimes\! \sigma^z_B  \; ;\nonumber \\
\hat H\!&\!\! =\!\! &\! \sum_i g_i \hat H_i \;,
\end{eqnarray}
where $\mathbf{1}_\alpha$ is the identity operator on the physical object $\alpha$ and the coefficient $g_j$ are linear combinations of $f(d_{\alpha,\beta})$.

Let us now consider that, at $t=0$, our system is in a state that is a tensor product of two states, each one of them defined on a single object.
Each one of these two states is a linear superposition $c_\alpha^{(1)}\ket{1_\alpha}+c_\alpha^{(2)}\ket{2_\alpha}$, $\alpha=A,B$.
For sake of simplicity we assume, in both cases, that the linear superposition is symmetric and real, i.e. 
$\ket{\psi_i(0)}=\frac{1}{\sqrt{2}}(\ket{1_i}+\ket{2_i})$ and  $\ket{\Psi(0)}=\ket{\psi_A(0)}\otimes\ket{\psi_B(0)}$.

The initial state  $\ket{\Psi(0)}$ evolves dynamically as
$\ket{\Psi(t)}=\hat U(t)\ket{\Psi(0)}$, where $\hat U(t)=\exp(-\imath \hat H t)$ is the time evolution operator and $\hat H$ is defined in Eq.~(\ref{Hamiltonian1}). 
Since the Hamiltonian is the sum of a commuting set of operators, $\hat U(t)$ can be factorized into 
the product of four unitary operators each one depending on one single term of the Hamiltonian, i.e. $\hat U(t)=\prod_i \hat U_i(t)$ where 
$\hat U_i(t)=\exp(-\imath g_i \hat H_i t)$.

It is easy to see that the rise of the entanglement in the system depends only on the action of $\hat U_4(t)$. 
In fact, $\hat U_1(t)$ is proportional to the identity operator and hence it contributes to the evolution only with a global phase factor.
On the other hand $\hat U_2(t)$ and $\hat U_3(t)$, being $\hat H_2= \sigma^z_A \otimes \mathbf{1}_B$ and $\hat H_3= \mathbf{1}_A \otimes \sigma^z_B$, act as 
local operators and hence are unable to create entanglement.
Therefore if entanglement between $A$ and $B$ is present it has to be related to the action of 
\mbox{$\hat U_4(t)=\exp(-\imath g_4 t \, \sigma^z_A \otimes \sigma^z_B )$}.
Hence, for what concerns the entanglement properties, the state $\ket{\Psi(t)}=\hat U(t)\ket{\Psi(0)}$ is equivalent to the state $\hat U_4(t) \ket{\Psi(0)}$ 
that is
\begin{eqnarray}
 \label{statet}
\!\!\!\!\hat U_4(t)\! \ket{\Psi\!(0)} \!\!&\!\!=\!\!&
\!\!\frac{e^{-\imath \phi}}{2}\!\left(e^{2 \imath \phi}\! \ket{1,\!1}\!+\!\! \ket{1,\!2}\!+\!\! \ket{2,\!1}
\!+\!e^{2 \imath \phi}\! \ket{2,\!2} \right), 
\end{eqnarray}
where the phase $\phi$ is given by
\begin{eqnarray}
 \label{phase}
\!\phi\!&\!=\!&\!\!\int_0^t \! g_4(\!\tau\!)  d\tau \\
\!\!&\!=\!&\!\!\int_0^t \!\left[f(d_{1,1}(\!\tau\!))\!+\!f(d_{2,2}(\!\tau\!))\!-\!f(d_{1,2}(\!\tau\!))\!-\!f(d_{2,1}(\!\tau\!))\right]\! d\tau. \nonumber
\end{eqnarray}
For a state described by Eq.~(\ref{statet}) the entanglement can be quantified by the concurrence $\mathcal{C}$~\cite{Hill1997,Wootters1998}. 
In this specific case it equals $\mathcal{C}=|\sin(2 \phi)|$.
However, in agreement with well-known quantum information results~\cite{Bennet1999,Nielsen2000}, and the conclusions of Ref.~\cite{Marletto2017} for a 
Koopman-type dynamics~\cite{Koopman1931}, an intrinsically classical interaction cannot generate entanglement.
Hence, in the case of a classic interaction, we must have that $\mathcal{C}=0$ for all possible set of the parameters of the system.
Since both the time-dependent distances and the integration time are arbitrary, to ensure that 
$\hat U_4(t) \ket{\Psi(0)}$ remains separable for any time $t>0$, we have to require that the function inside the integral in Eq.~(\ref{phase}) must vanish 
for all possible sets   of distances, i.e.
\begin{equation}
\label{condition}
f(d_{1,1})+f(d_{2,2})-f(d_{1,2})-f(d_{2,1})=0.
\end{equation}
All the interactions for which at least one physically achievable setup exists that does not satisfy Eq.~(\ref{condition}) can generate entanglement.
Then, a formulation of the fundamental theory based on a classical mediating field leading to the {\it effective} nonlocal interaction $f(d_{\alpha,\beta})$
would prove to be inconsistent with entanglement generation.
Eq.~(\ref{condition}) was derived within the hypothesis that the interaction depends only on the distance ad not on a generic function of the relative position 
of the states of the two objects. 
The extension to more general functions is straightforward.

We are now in the position to prove that eq.~(\ref{condition}) is violated by all non-trivial conservative interactions.
Let us define the vector $\mathbf{x}$ in $\mathbb{R}^{\otimes12}$ that denotes the four coordinates in real space of the internal states in Fig.~\ref{figure_1} 
and the function $h(\mathbf{x})=f(d_{1,1})+f(d_{2,2})-f(d_{1,2})-f(d_{2,1})$.
Eq.~(\ref{condition}) can then be rewritten as $h(\mathbf{x})=0$.
Such condition, as the interaction is conservative, 
expresses a holonomic constraint of the corresponding dynamics, 
as it depends just on the vector of the coordinates but not on the velocities or any 
higher order derivative with respect to time.
We name $\mathcal{L}$ the set of points $\mathbf{x} \in \mathbb{R}^{\otimes12}$ that satisfy the holonomic constraint. 
It is immediate to note that the trivial interaction $f(d)=const$ fulfills the constraint for all possible configurations and hence 
$\mathcal{L} \equiv \mathbb{R}^{\otimes12}$.
Besides, for a generic interaction one has $\mathcal{L}$ is not an empty set. 
For example, the subset of configurations which fulfill $d_{1,1}=d_{1,2}$ and $d_{2,1}=d_{2,2}$ always satisfies $h(\mathbf{x})=0$. 
However, for a generic position dependent interaction, we have that $\mathcal{L}\not\equiv \mathbb{R}^{\otimes 12}$.
Indeed, let us consider $\mathbf{x}_0\in\mathcal{L}$ and $\mathbf{\hat{n}} \delta$ a vector of modulus $\delta$ and direction $\mathbf{\hat{n}}$ defined in 
${R}^{\otimes 12}$. 
If the function $f(d)$ is analytic also $h(\mathbf{x})$ will be analytic and assuming $\delta$ small enough we can write 
$h(\mathbf{x}_0+\mathbf{\hat{n}}\delta)=\frac{\partial h}{\partial \mathbf{\hat{n}}}|_{\mathbf{x}=\mathbf{x}_0} \delta$.
But if $\mathcal{L} \equiv \mathbb{R}^{\otimes 12}$, also $\mathbf{x}_0+\mathbf{\hat{n}}\delta$ should lay in $\mathcal{L}$ and, since $\delta \neq 0$ we must 
have $\frac{\partial h}{\partial \mathbf{\hat{n}}}|_{\mathbf{x}=\mathbf{x}_0}=0$ for all $\mathbf{\hat{n}}$ and all $\mathbf{x}_0$.
This result can be generalized to derivatives of all orders and, therefore, to have that $\mathcal{L}\equiv \mathbb{R}^{\otimes 12}$ the interaction potential $f(d)$ must be constant all over the space. 
We then conclude that a non-trivial interaction induces entanglement whenever 
$\mathbf{x} \not\in\mathcal{L}$.

As an example we consider an interparticle potential $f(d)=\lambda\, d^{-\alpha}$ where $\lambda$ is the coupling constant and $\alpha$ the 
power-law exponent of the decay of the interaction with distance. 
For the gravitational potential in the limit of weak field and non-relativistic velocities $\lambda_G=-G m_A m_B$ while, for the Coulomb potential 
$\lambda_C=q_A q_B/{4\pi\varepsilon_0}$, where $m_i$ and $q_i$ are, respectively the masses and the electric charges of the two objects.
In both cases $\alpha=1$. 
Analogously, realistic atomic, ionic, or molecular potentials may be cast in a similar form with proper coupling constants and power-law exponents.
For sake of simplicity we choose:
1) the two quantum objects $A$ and $B$ at rest in the same inertial reference system;
2) the fours states on the same line; 
3) the distance between the two states of the same object equal to be $\delta_x$. 
Accordingly with these choice we have that the four distances are $d_{1,1}=x$, $d_{2,1}=d_{1,2}=x+\delta_x$, $d_{2,2}=x+2 \delta_x$ and$\delta_x\ll x$.
With this assumption we have that the phase $\phi(t)$ varies linearly with $t$ and it is given by 
\begin{equation}
\label{conditiongrav}
\phi(t)= \lambda \, t \frac{\alpha(\alpha+1) \, \delta_x^2}{x^{2+\alpha}} \; .
\end{equation}
The phase in Eq.~(\ref{conditiongrav}) vanishes only in the trivial case in which $\delta_x=0$, i.e. if the distance between the two states of the same physical
objects goes to zero and the interaction does not depend on the state anymore.

An experimental test of what we have proven so far can be realized using
levitated diamond nanocrystals with nitrogen-vacancy (NV) centers in high 
vacuum as discussed in~\cite{HSu2016, Bose2017}. 
These proposals focussed on the coherent superposition of spatially separated states of
massive objects to probe the quantization of the gravitational field.
However, as mentioned above, the basic requirement is to generate a state dependent interaction.
This can be achieved even in the absence of spatial separation of the internal states.
For example one can consider particles in which the internal states are characterized by different masses,
as neutrinos~\cite{Marletto18}.
Alternatively one can take into account Rydberg atoms in microtraps which can be optically manipulated
with high precision~\cite{Pillet16}. 
For such system the two internal states would correspond to either the ground state and one highly
excited Rydberg level or two excited states interacting electromagnetically~\cite{Ravets14,Labuhn16,Jau16}.

In general, when studying the interaction between two bodies, it is often useful to describe 
their effective interaction potential 
in terms of a multipole expansion.
Each multipole of the series contains an angular dependence as well as an inverse power of the distance among
a (composite) body and a reference point or another body.
This standard procedure is widely used in problems involving gravitational systems of masses or
electric and magnetic distributions of charges and currents.
The calculation and measurement of multipole moments to characterize the interaction potential 
is relevant to the experimental setups mentioned above.
To this purpose, we propose to make use of the analysis of the entanglement to 
estimate the strength of the subdominant terms of the potential. 

\begin{figure}[t]
\begin{center}
\includegraphics[width=1.0\columnwidth]{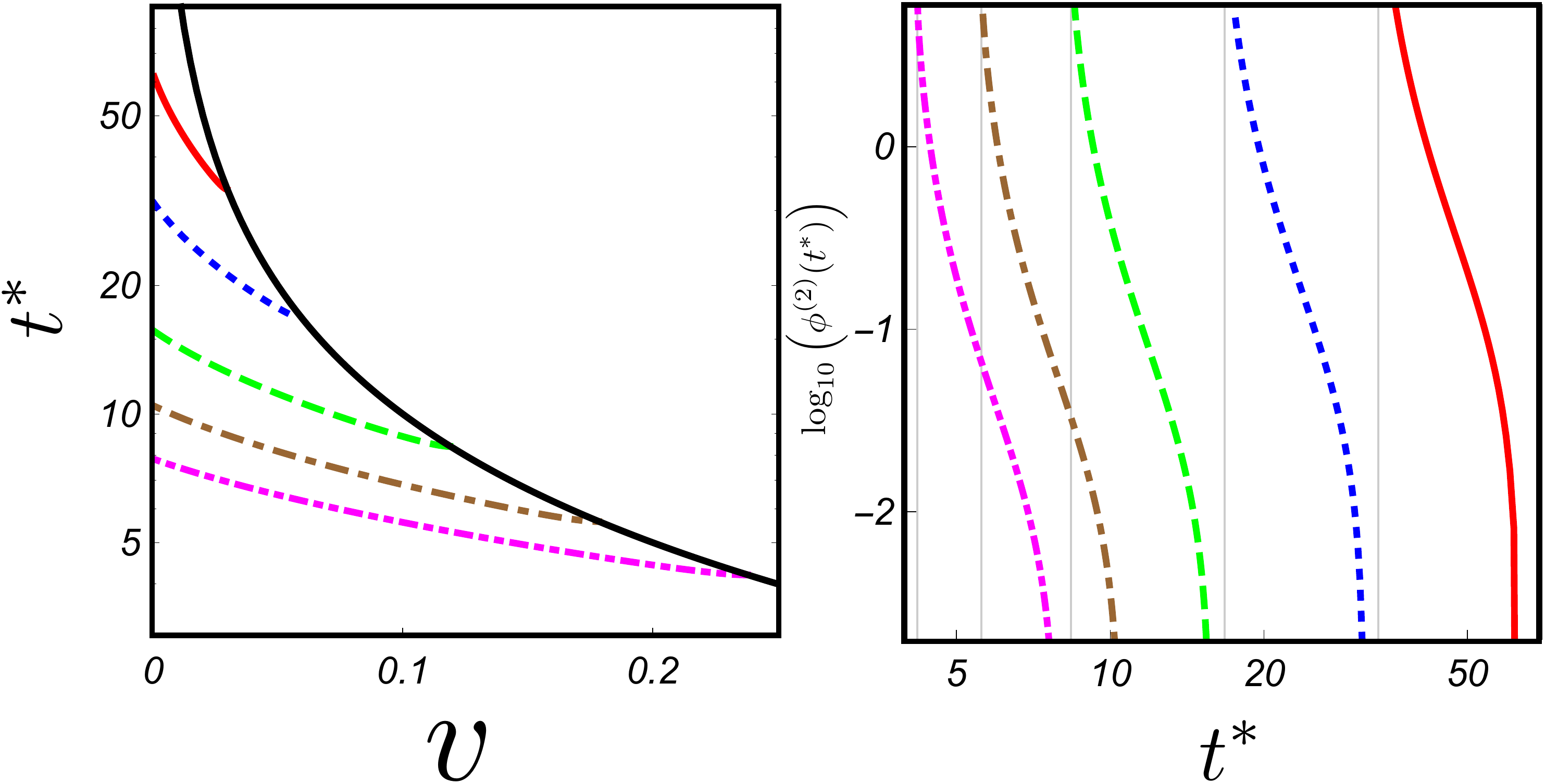}
\caption{Optimal time $t^*$ and dynamical phase $\phi^{(2)}(t^*)$. 
{\it (a):} $t^*$ as function of the velocity $v$ for different values of $\omega$.
{\it (b):} $\log_{10}\phi^{(2)}(t^*)$ for the $t^*$ evaluated in the left plot for several value of $\omega$.
For both plots we set: $\omega=0.05$ (red solid); $\omega=0.1$ (blue dotted line); $\omega=0.2$ (green dashed line);
$\omega=0.3$ (brown dot-dashed line); $\omega=0.4$ (magenta dot-dot-dashed line).
The black line in the left figure displays the time $\bar{t}=L/v$ 
in which the two physical objects collide.
The gray vertical grid lines in the right plots indicate, for each $\omega$ the time at which $t^*=\bar{t}$.
In the numerical simulation we assumed $\delta x_0=0.1\, L$ and we normalized all quantities in such a way that $c_2=1$.
}
\label{figure_3}
\end{center}
\end{figure}

As an example we consider a Laurent expansion for the function $f(d)$, neglecting for simplicity angular dependence of the multipoles,
\begin{equation}
\label{laurent}
f(d) = \sum_{n} f_n(d)= \sum_{n} c_n d^{-n}
\end{equation}
where each term has a definite power-law dependence on distance~\cite{footnote1}. 
To increase the signal-to-noise ratio we design a protocol in a way that the 
entanglement generated by the main terms vanishes.
Focusing on the $n$-th term of the potential, assuming \mbox{$|f^{(n)}\!(d)|\!\gg\! |f^{(n+1)}\!(d)|$}, and defining 
\begin{eqnarray} \label{constraints_1}
\!\!\!\!\!\!\phi^{(i)}\!\!&\!\!=\!\!& c_n\!\!\int_0^t \!\!\left[\frac{1}{d_{1,1}(t)^n\!}\!+\!\frac{1}{d_{2,2}(t)^n\!}\!-\!\frac{1}{d_{1,2}(t)^n\!}
\!-\!\frac{1}{d_{2,1}(t)^n\!}\right]\! d\tau, 
\end{eqnarray}
we have to set $\sum_{i=1}^{n-1} \phi^{(i)}=0$.
We apply this scheme to the case of the first correction to a potential whose leading term is $f^{(1)}(d)=c_1\,d^{-1}$. The first subleading term then reads 
$f^{(2)}(d)=c_2\,d^{-2}$.
\begin{figure}[t!]
\begin{center}
\includegraphics[width=1.0\columnwidth]{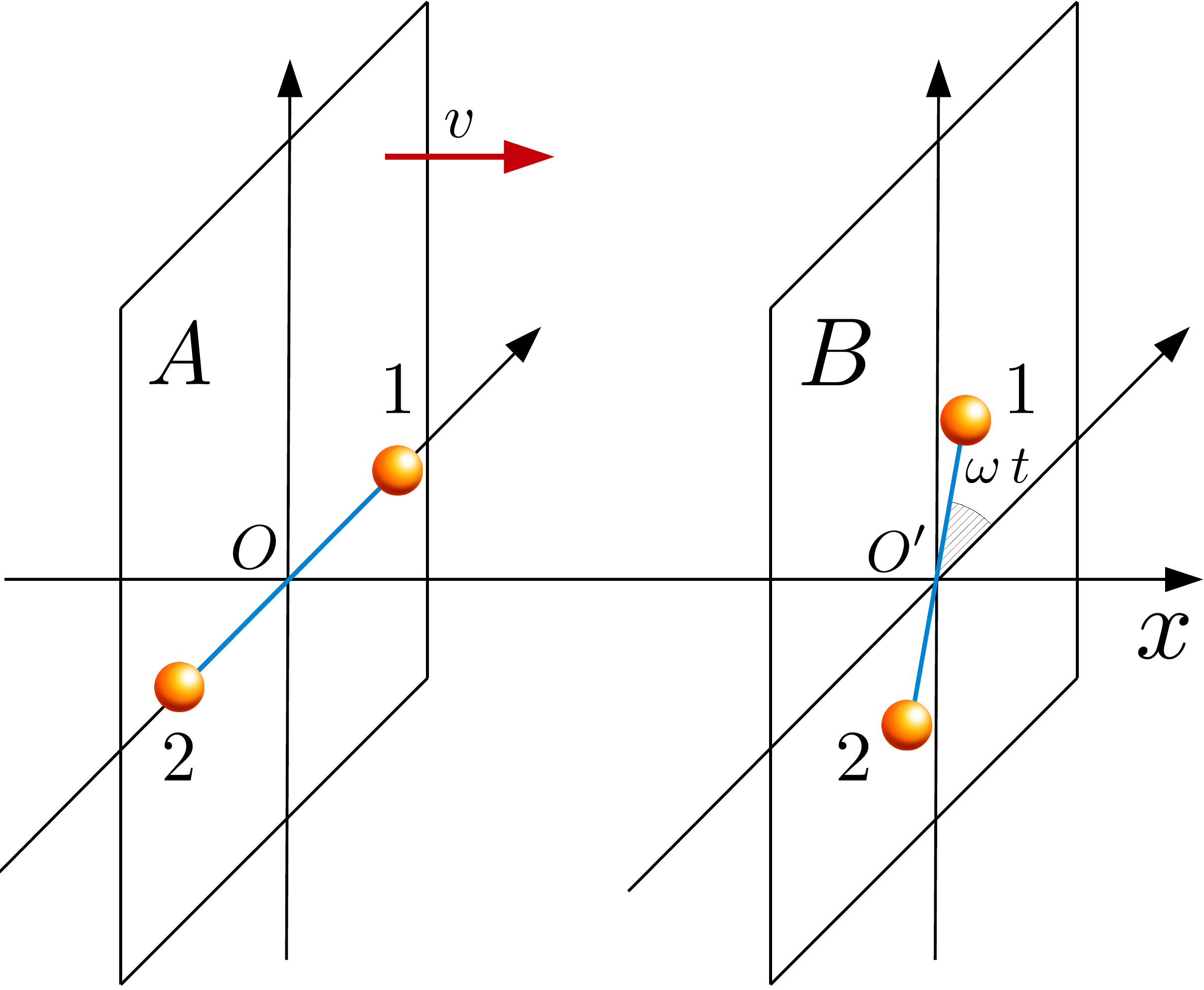}
\caption{Sketch of the proposed experimental protocol.
The two objects lie on two parallel planes whose relative distance varies in time with a constant velocity $v$.
The object $A$ is at rest while $B$ rotates with an uniform angular velocity $\omega$ around the center of the segment joining the positions associated with 
its two internal states.}
\label{figure_2}
\end{center}
\end{figure}
We then consider a system as in Fig.~(\ref{figure_2}):
1) Two identical objects lie on two parallel planes;
2) The distance between the two internal states of each object is $x_0$;
3) The line joining the midpoints of the center of mass of each body is perpendicular to the planes;    
4) The inter-plane distance reduces at constant rate $v$, i.e. it follows the law $L-vt$;
5) The object $B$ rotates on its plane with uniform angular velocity $\omega$ around the midpoint of the segment passing through the two states.
Then, the state dependent distances become
\begin{eqnarray}
d_{11}(t) = & \! d_{22}(t) = &\! 
\left[ (L-vt )^2+x_0^2\sin^2(\omega t/2) \right]^{1/2}, \nonumber \\
d_{12}(t) = & \! d_{21}(t) =  &\! 
\left[ (L-vt )^2+x_0^2\cos^2(\omega t/2) \right]^{1/2}.
\end{eqnarray}
Said $t^*$ the optimal time at which $\phi^{(1)}=0$ we have that, if $v,\omega\neq0$  then $\phi^{(2)}(t) \neq 0$.
In the left panel of Fig.~\ref{figure_3} we plot $t^*$ as function of $v$ for different $\omega$ whereas in the right one we show the behavior of the phase
$\phi^{(2)}(t^*)$ induced by the sub-leading term $\propto d^{-2}$.
As general result we can see that $\phi^{(2)}(t^*)$  increases decreasing $v$ and/or $\omega$  and tends to diverge when $t^*$ goes towards $\bar{t}=L/v$, 
i.e. when the time in which $\phi^{(1)}$ vanishes coincides with the time at which the two physical objects collide.
In fact, bringing $t^*$ versus $\bar{t}$ and reducing $v$ and $\omega$ we maximize the total integration time and we enter in a region of the space in which
the subleading term is more relevant. 
Notably this protocol allows to measure the dynamical phase in a much 
faster time interval than the one would get upon keeping interparticle distances 
fixed.
This is of utmost relevance for any possible experimental realization. 
Indeed decoherence effects or losses reduce, over time, 
the coherent superposition in a statistical mixture in which all entanglements 
vanish  in any realistic device~\cite{Brune1996,Paganelli2002,Hornberger2003}.
We notice that this protocol applies well to both gravitational and electromagnetic interactions, i.e. to mass or charge/current distributions. Rotations of internal states can be performed by using external magnetic (e.g. in diamonds nanocrystals) or electric fields (Rydberg atoms) in the cited experimental setups~\cite{footnote2}.

In this letter we proved that for every non-constant potential
generating entanglement between two particles, one has to associate a quantized 
description of such interaction.
The proof is based on a holonomic constraint that must be satisfied at any time 
$t$ if the mediating field were classical for it not to generate entanglement. 
This requirement, however, would rule out all non-trivial interactions.  
The scheme discussed in the first part can be easily generalized to 
include more complex dependence of the potential on the relative distance or 
state-dependent interactions.
We have also shown how it is possible 
to measure the various terms of a multipole-like Laurent expansion of an interaction potential
via the measurement of the entanglement between the two objects.
This scheme is based on a proper manipulation of spatial and internal degrees
of freedom of a two-body system during the interaction time.
From a fundamental perspective, the results presented in this letter were 
obtained in the non-relativistic limit and for a flat space-time.
Going beyond these limitations would not alter the general picture that 
we presented here, and they will be considered in future works.

\begin{acknowledgments}
{\it Acknowledgments}. 
We acknowledge discussions with C. Bonato and F. Novaes.
S.M.G. acknowledge support by the H2020 CSA Twinning project No. 692194, "RBI-T-WINNING''. 
T.M. acknowledges for support CAPES through the Project CAPES/Nuffic n. 88887.156521/2017-00 
and CNPq through Bolsa de produtividade em Pesquisa n. 311079/2015-6.
\end{acknowledgments}


\begin{thebibliography}{99}
\bibitem{footnote1}
{Also in this case a straightforward generalization to interaction potentials depending to the relative coordinates
and not only on the relative distance can be easily done.}
\bibitem{footnote2}
Post-Newtonian and quantum corrections for the gravitational interaction between two mesoscopic objects can also be written as an expansion of the form of eq.(\ref{laurent}) for two point-like bodies. However, the application of this scheme for such terms 
would not produce observable effects.
Indeed the relativistic corrections are proportional to $\ell_{cl}/r$, where $\ell_{cl}=Gm/c^2=7.4\cdot 10^{-42}$m for a mass $m=10^{-14}$kg. Quantum corrections are 
proportional to $(\ell_P/r)^2$, where $\ell_P=1.6\cdot 10^{-35}$m is the Planck length.
\end{thebibliography}
\end{document}